\documentclass{article}
\usepackage{graphics}
\usepackage{rotating}
\usepackage{emulateapj}

\makeatletter
\let\@internalcite\cite
\def\cite{\def\astroncite##1##2{##1\ ##2}\@internalcite}
\def\citey{\def\astroncite##1##2{##1\ (##2)}\@internalcite}
\def\citeyear{\def\astroncite##1##2{##2}\@internalcite}
\def\@citex[#1]#2{\if@filesw\immediate\write\@auxout{\string\citation{#2}}\fi
  \def\@citea{}\@cite{\@for\@citeb:=#2\do
    {\@citea\def\@citea{; }\@ifundefined
       {b@\@citeb}{{\bf ??}\@warning
       {Citation `\@citeb' on page \thepage \space undefined}}%
{\csname b@\@citeb\endcsname}}}{#1}}
\def\@cite#1#2{#1\if@tempswa #2\fi}
\def\@biblabel#1{}
\def\astroncite#1#2{#1\ #2}
\makeatother

\newcommand{\aql}{V1408~Aql}

\begin{document}

\slugcomment{Accepted for Publication in The Astrophysical Journal}

\lefthead{Nowak \& Wilms}
\righthead{V1408 Aql (4U1957+11)}

\title{On the enigmatic X-ray Source V1408~Aql (=4U~1957+11)}

\author{Michael A. Nowak\altaffilmark{1} and J\"orn Wilms\altaffilmark{2}}

\altaffiltext{1}{JILA, University of
  Colorado, Campus Box 440, Boulder, CO~80309-0440, USA;
  mnowak@rocinante.colorado.edu}  
\altaffiltext{2}{Institut f\"ur Astronomie und Astrophysik,
  Abt.~Astronomie, Waldh\"auser Stra\ss{}e 64, D-72076 T\"ubingen, Germany;
  wilms@astro.uni-tuebingen.de} 

\received{1998 December 4}
\accepted{1999 March 11}

\begin{abstract}
  
  Models for the characteristically soft X-ray spectrum of the compact
  X-ray source \aql\ (=4U\,1957+11) have ranged from optically thick
  Comptonization to multicolor accretion disk models.  We critically
  examine the X-ray spectrum of \aql\ via archival Advanced Satellite for
  Cosmology and Astrophysics (ASCA) data, archival R\"ontgensatellit
  (ROSAT) data, and recent Rossi X-Ray Timing Explorer (RXTE) data.
  Although we are able to fit a variety of X-ray spectral models to these
  data, we favor an interpretation of the X-ray spectrum as being due to an
  accretion disk viewed at large inclination angles.  Evidence for this
  hypothesis includes long term (117\,day, 235\,day, 352\,day)
  periodicities seen by the RXTE All Sky Monitor (ASM), which we interpret
  as being due to a warped precessing disk, and a 1\,keV feature in the
  ASCA data, which we interpret as being the blend of L fluorescence
  features from a disk atmosphere or wind.  We also present timing analysis
  of the RXTE data and find upper limits of 4\% for the root mean square
  (rms) variability between f=$10^{-3}-16$\,Hz.  The situation of whether
  the compact object is a black hole or neutron star is still ambiguous;
  however, it now seems more likely that an X-ray emitting, warped
  accretion disk is an important component of this system.

\end{abstract}


\setcounter{footnote}{0}

\section{Introduction}\label{sec:intro}
The low mass X-ray binary (LMXB) \aql\ ($=$4U\,1957+11, 3U\,1956+11) was
detected during scans of the Aquila region by \textsl{Uhuru} in 1973
(\cite{giacconi:74a}), and it was subsequently identified with an 18\fm{}7
star having a strong blue excess (\cite{margon:78a}). The object is
situated in a region of relatively small extinction ($N_{\rm H}\approx
1.3\times 10^{21}\,{\rm cm^{-2}}$; \cite{dickey:90a,stark:92a}). $A_{\rm
  V}$ measurements place the source at a distance $>$2.5\,kpc, and
comparisons of its X-ray and optical luminosity to Sco~X-1 place it at a
distance of $\sim$7\,kpc (\cite{margon:78a}).

Little is known about the nature of the system. Optical spectra of \aql\ 
reveal a power-law continuum with H$\alpha$, H$\beta$, and He\,{\sc ii}
4686\AA\ emission lines (\cite{cowley:88a,shahbaz:96a}), typical for an
accretion disk-dominated system. \citey{thorstensen:87a} reported a nearly
perfectly sinusoidal V-band luminosity modulation with 10\% amplitude and a
0.389\,d (=9.33\,h) period, which he interpreted as due to X-ray heating of
the companion. In recent multicolor photometry a more complex lightcurve
with 30\% modulation amplitude was observed. \citey{hakala:99a} interpret
this change in the shape of the lightcurve as evidence for a disk with a
large outer rim, possibly due to a warped disk, seen close to edge on
(\cite{hakala:99a}; see also \S\ref{sec:long}). This interpretation is also
consistent with the shape of the infrared spectrum (\cite{smith:90a}).  The
short orbital period is indicative of a late type main sequence star of $M
\sim 1\,{\rm M}_\odot$ as the donor star. The absence of X-ray eclipses and
the assumption that the donor star fills its Roche lobe yield an upper
limit on the orbital inclination of $i\sim 70^\circ$--$75^\circ$,
consistent with the models for the optical variability (\cite{hakala:99a}).

\aql\ is one of the less well-studied possible black hole candidates
(BHCs).  Identification as either a BHC or a neutron star-low mass X-ray
binary (NS-LMXB) is usually made by analogy with the spectral- and
timing-behavior of better observed sources.  \aql\ has been a BHC since
1984, when EXOSAT X-ray observations revealed that \aql\ has a very soft
X-ray spectrum, similar to that of other BHC.  In color-color-diagrams,
\aql\ lies halfway between the black hole candidate GX~339$-$4 (in its high/soft
state) and the neutron-star LMXBs Cyg~X-2 and LMC~X-2
(\cite{white:84a,schulz:89a}).  This color identification of \aql\ as a
BHC, however, is not definitive.

Previous narrow-band observations have not characterized the X-ray spectrum
in a consistent manner.  The analysis of 1983 and 1985 EXOSAT observations
of \aql\ led to contradictory results. While \citey{singh:94a} succeeded in
fitting a Comptonization spectrum to these data and interpreted this as an
indication that \aql\ is a black hole candidate, \citey{ricci:95a}
interpreted the same data as being similar to that observed from NS-LMXBs.
Observations with \textsl{Ginga}, with its larger spectral range and
effective area, have shed more light on the nature of \aql\ 
(\cite{yaqoob:93b}).  The values of the normalizations of multicolor disk
models (MCD; \cite{mitsuda:84a}), i.e. $(r_{\rm in}/d)^2 \cos i$ where
$r_{\rm in}$ is the inner disk radius, $d$ is the distance to the source,
and $i$ is the inclination, have been used to distinguish between BHCs and
NS-LMXBs (\cite{tanaka:95a}).  In the case of \aql, $r_{\rm in} \cos^{1/2}
i \approx 2~{\rm km}$ assuming $d = 7$\,kpc, which is more characteristic
of sources containing neutron stars. Additionally, the \textsl{Ginga}
observation showed evidence of a hard tail (1--18\,keV) comprising $\sim
25\%$ of the inferred flux for this system at that time.  The best fit
power-law photon indices for the hard component ranged from $\Gamma\approx
2$ to $3$.

The EXOSAT observations of \citey{ricci:95a} indicate the presence of an
iron fluorescence line with an equivalent width of 90\,eV or smaller and a
line-energy of 7.06\,keV (i.e., highly ionized). Other values in the
literature range from non-detection (e.g., \cite{yaqoob:93b}) to 200\,eV
(\cite{white:84a}), the uncertainty being mainly due to the difference in
the assumed spectral continua and the different sensitivities of the
instruments. 

Except for one observation, which hints toward a weak red-noise
($f^{-\alpha}$) component between $10^{-4}$ and $10^{-3}$\,Hz, all EXOSAT
observations are consistent with the absence of any periodic features
(\cite{ricci:95a}).  The \textsl{Ginga} observations have yet to have their
short timescale variability analyzed; however, they do show evidence of
significant flux and color changes on long time scales ($\gtrsim 10^4$\,s).
The \textsl{Vela 5B} satellite did not detect any long-term X-ray variability
from this source (\cite{priedhorsky:84a}); however, the upper limits to the
variability were not particularly strong.

If the published spectral models are accepted at face value, then the {\it
  relative} energetics of the disk black-body and power-law components, as
well as the slope of the high-energy power-law, are very similar to those
seen in BHCs such as \mbox{LMC~X-1} (\cite{ebisawa:89a,wilms:98d}),
\mbox{LMC~X-3} (\cite{treves:88a,wilms:98d}), and in the soft state of
\mbox{GX~339$-$4} (\cite{miyamoto:91a,grebenev:91a}).  However, at
luminosities as low as that of \aql, BHCs tend to show hard tails with no
evidence of a disk or thermal component.  On the other hand, NS-LMXBs that
exhibit soft disk spectra also tend to show an additional $\sim 2$\,keV
blackbody component, while showing little hard flux (\cite{miyamoto:94b},
and references therein). Furthermore, low-luminosity NS-LMXBs \emph{also}
tend to be dominated by hard emission.

Thus, there are good arguments that point towards \aql\ being a neutron
star and also toward it being a black hole; however, none of the arguments
are truly conclusive. In either case, \aql\ would still be a unique object,
being either an unusually soft low-luminosity BHC, an unusually soft
low-luminosity neutron star, or a soft neutron star with an unusually
energetic hard tail. With the advent of X-ray detectors with much larger
effective areas than EXOSAT and \textsl{Ginga}, as well as with the
availability of detectors of higher energy resolution, such as those on the
Advanced Satellite for Cosmology and Astrophysics (ASCA), a critical
reexamination of the X-ray spectrum of \aql\ has become possible. In this
paper we present the results from our analysis of a 30\,ksec pointed
observation with the Rossi X-Ray Timing Explorer (RXTE), as well as
archival data from ASCA and the R\"ontgensatellit (ROSAT). In
\S\ref{sec:spectral} we present the results from the spectral analysis. We
discuss the timing analysis in \S\ref{sec:temporal}, the long term
variability of the source in \S\ref{sec:long}, and we discuss our results
in \S\ref{sec:disc}. The details of the data extraction are described in an
appendix.

\begin{table*}
\begin{center}
\caption{\small Observing log for \aql. Count rates given are background
  subtracted. 
\label{tab:obslog}}
\begin{tabular}{llll}
\hline
\hline
Date                         &  On-Source Time   & Instrument  & Count Rate \\
                             & [ksec]            &             & [cps]      \\
\hline
1997 November 26.02--26.45 & 21.2        & RXTE-PCA    & 230 \\
1997 November 27.08--27.17 & ~4.7        &             & 229 \\
1997 November 29.08--29.15 & ~1.5        &             & 233 \\
\hline
1994 October 31.13--31.93  & 21.6        & ASCA SIS0   & 9.8 \\
                           &             & ASCA SIS1   & 7.0 \\
                           &             & ASCA GIS2   & 19.4 \\
                           &             & ASCA GIS3   & 21.5 \\
\hline
1992 May 08.39--11.14      & 14.6        & ROSAT PSPC  & 19.2 \\
\hline
\end{tabular}
\end{center}
\end{table*}

\section{Spectral Analysis}\label{sec:spectral}
\aql\ was observed with RXTE in 1997 November in three observing
blocks for a total on source time of 27\,ksec. A log of the
observations is given in Table~\ref{tab:obslog}. Since the spectral
shapes of the three observing blocks are identical, the data were
analyzed together. Spectral and temporal data were extracted using the
methods outlined in appendix~\ref{sec:rxte}. Spectral analysis was
performed with XSPEC, version 10.0s (\cite{arnaud:96a}).

The RXTE spectrum of \aql\ is very soft. The Proportional Counter Array
(PCA) did not detect any flux above $\sim$20\,keV and the High Energy X-ray
Timing Experiment (HEXTE) count rates are consistent with zero: the
background subtracted count rates were $0.4\pm 0.2$\,cps and $0.0\pm
0.2$\,cps, for HEXTE clusters~A and~B, respectively. The residual flux in
cluster~A is most probably due to a slight overestimation of the HEXTE
background dead time, as the spectrum seen is similar to the HEXTE
background. Therefore, we do not consider \aql\ to be detected with HEXTE
and will not further discuss these data.


\newcounter{mocnt}\setcounter{mocnt}{0}
\renewcommand{\themocnt}{\ensuremath{{}^\alph{mocnt}}}
\newcommand{\molab}[1]{\refstepcounter{mocnt}#1\tablenotemark{\themocnt}\label{#1}}
\newcommand{\modes}[2]{\tablenotetext{\ref{#1}}{#2}}

\begin{deluxetable}{llllll}
\tablecaption{Results of spectral fitting\label{tab:fitres}}
\tablehead{
\colhead{Model} & \colhead{Detector} &  \multicolumn{3}{c}{Parameters} 
                                                            &  \colhead{$\chi^2/{\rm dof}$}
}
\startdata
\small \molab{cutoffpl}
          & \small  PCA 
                & \small  $\Gamma=-0.26\pm0.06$ 
                & \small  $A_{\rm PL}=0.21\pm0.01$
                & \small  $E_{\rm fold}=1.44\pm0.02$
                & \small  60.2/30  \nl
          & \small  SIS 
                & \small  $\Gamma=0.43\pm0.04$ 
                & \small  $A_{\rm PL}=0.200\pm 0.002$
                & \small  $E_{\rm fold}=2.30\pm0.07$
                & \small   559/501 \nl
          & \small      
                & \small   $N_{\rm H}=1.4\pm0.01$
                & \small
                & \small          \nl 
          & \small  GIS 
                & \small  $\Gamma=0.56\pm0.02$ 
                & \small  $A_{\rm PL}=0.267\pm 0.001$
                & \small  $E_{\rm fold}=2.44\pm0.03$
                & \small  1630/1342\nl
\small \molab{power}
          & \small PSPC
                & \small  $\Gamma=1.54\pm0.05$ 
                & \small  $A_{\rm PL}=0.148\pm 0.004$
                & \small  $N_{\rm H}=1.8\pm0.1$
                & \small   56.7/21 \nl
\hline
\small \molab{diskbb}
          & \small  PCA 
                & \small  $kT_{\rm in}=1.319\pm 0.003$
                & \small  $A_{\rm diskbb}=15.7\pm 0.26$ 
                & \small 
                & \small  42.0/31  \nl 
          & \small  SIS 
                & \small  $kT_{\rm in}=1.530\pm 0.001$
                & \small  $A_{\rm diskbb}=8.7\pm 0.2$ 
                & \small  $N_{\rm H}=1.32\pm0.04$
                & \small  628/502  \nl
          & \small  GIS 
                & \small  $kT_{\rm in}=1.487\pm 0.004$
                & \small  $A_{\rm diskbb}=12.2\pm 0.1$ 
                & \small 
                & \small  2148/1343\nl
          & \small  PSPC
                & \small $kT_{\rm in}=1.0$ 
                & \small  $A_{\rm diskbb}=30.3$ 
                & \small  $N_{\rm H}=1.0$
                & \small   104/22  \nl
\hline
\small \molab{compTT}
          & \small  PCA 
                & \small  $kT_0=0.338^{+0.08}_{-0.337}$
                & \small  $kT_{\rm e}=1.17\pm 0.01$
                & \small  $\tau=10.7\pm 0.3$ 
                & \small  23.8/29  \nl
          & \small      
                & \small  ${\rm norm}=0.33^{+2.5}_{-0.05}$
                & \small 
                & \small 
                & \small      \nl 
          & \small  SIS 
                & \small  $kT_0=0.32\pm 0.01$          
                & \small  $kT_{\rm e}=1.28^{+0.03}_{-0.02}$       
                & \small  $\tau=12.2^{+0.2}_{-0.1}$ 
                & \small  580/500  \nl
          & \small      
                & \small  ${\rm norm}=0.239\pm 0.003$
                & \small  $N_{\rm H}=0.37^{+0.16}_{-0.14}$
                & \small 
                & \small           \nl
          & \small  GIS 
                & \small  $kT_0=0.27\pm 0.01$ 
                & \small  $kT_{\rm e}=1.24\pm 0.01$      
                & \small  $\tau=12.9\pm 0.2$ 
                & \small  1371/1341 \nl
          & \small      
                & \small  ${\rm norm}=0.333\pm 0.005$
                & \small 
                & \small 
                & \small           \nl
          & \small PSPC
                & \small  $kT_0=0.176\pm 0.030$ 
                & \small  $kT_{\rm e}=2.40^{+20.}_{-1.0}$
                & \small  $\tau=9.8^{+5.5}_{-3.2}$
                & \small  29.5/29  \nl
          & \small      
                & \small   ${\rm norm}=0.15^{+0.11}_{-0.07}$
                & \small  $N_{\rm H}=1.07\pm 0.30$
                & \small             
                & \small           \nl
\hline
\small cutoffpl+ray
          & \small  PCA
                & \small  $\Gamma=-0.69^{+0.25}_{-0.19}$
                & \small  $A_{\rm PL}=0.12^{+0.01}_{-0.02}$
                & \small  $E_{\rm fold}=1.35\pm0.06$ 
                & \small   17.4/28 \nl
          & \small
                & \small  $kT_{\rm ray}=1.17\pm 0.19$
                & \small  $A_{\rm ray}=44^{+23}_{-20}$
                & \small  
                & \small           \nl 
          & \small  SIS
                & \small  $\Gamma=0.36\pm0.05$
                & \small  $A_{\rm PL}=0.196\pm 0.003$
                & \small  $E_{\rm fold}=2.21\pm0.07$ 
                & \small   507/499 \nl
          & \small
                & \small  $kT_{\rm ray}=1.08^{+0.06}_{-0.08}$
                & \small  $A_{\rm ray}=6.7^{+1.7}_{-1.8}$
                & \small  $N_{\rm H}=1.4\pm0.01$
                & \small           \nl 
          & \small   GIS
                & \small  $\Gamma=0.35\pm0.03$
                & \small  $A_{\rm PL}=0.251\pm 0.002$
                & \small  $E_{\rm fold}=2.16\pm0.04$
                & \small  1303/1340\nl
          & \small
                & \small  $kT_{\rm ray}=0.55^{+0.02}_{-0.09}$
                & \small  $A_{\rm ray}=32^{+9}_{-3}$
                & \small 
                & \small           \nl
\small power+ray 
          & \small PSPC
                & \small  $\Gamma=1.2\pm0.1$ 
                & \small  $A_{\rm PL}=0.12\pm 0.01$
                & \small  $N_{\rm H}=1.3\pm0.02$
                & \small   19.1/20  \nl
          & \small
                & \small  $kT_{\rm ray}=1.09^{+0.06}_{-0.09}$
                & \small  $A_{\rm ray}=12.6^{+4.8}_{-5.2}$
                & \small 
                & \small            \nl
\hline
\small
diskbb+\molab{ray}
          & \small  PCA
                & \small  $kT_{\rm in}=1.33\pm 0.01$ 
                & \small  $A_{\rm diskbb}=14.2^{+1.0}_{-0.7}$
                & \small  $kT_{\rm ray}=1.68^{+0.65}_{-0.19}$
                & \small   17.3/29 \nl
          & \small      
                & \small  $A_{\rm ray}=100^{+56}_{-48}$
                & \small
                & \small               
                & \small           \nl
          & \small  SIS 
                & \small  $kT_{\rm in}=1.55\pm 0.01$ 
                & \small  $A_{\rm diskbb}=8.3\pm 0.2$
                & \small  $N_{\rm H}=1.46\pm0.06$
                & \small   542/500 \nl
          & \small      
                & \small  $A_{\rm ray}=8.6^{+3.6}_{-1.5}$
                & \small  $kT_{\rm ray}=1.13^{+0.16}_{-0.03}$
                & \small               
                & \small           \nl
          & \small  GIS 
                & \small  $kT_{\rm in}=1.58 \pm 0.01$
                & \small  $A_{\rm diskbb}=9.14 \pm 0.26$
                & \small 
                & \small  1276/1341\nl
          & \small      
                & \small  $A_{\rm ray}=67^{+9}_{-8}$
                & \small  $kT_{\rm ray}=1.64^{+0.14}_{-0.19}$
                & \small           \nl
          & \small PSPC
                & \small  $kT_{\rm in}=1.43^{+0.24}_{-0.15}$
                & \small  $A_{\rm diskbb}=8.5\pm 3.6$            
                & \small  $N_{\rm H}=0.8\pm 0.1$
                & \small  23.2/20  \nl
          & \small      
                & \small  $A_{\rm ray}=12.7^{+10}_{-4}$
                & \small  $kT_{\rm ray}=1.1\pm 0.1$
                & \small 
                & \small           \nl
\hline
\enddata
\footnotesize

\modes{cutoffpl}{Exponentially cut-off power-law of the form $A_{\rm PL}
          E^{-\Gamma} \exp(-E/E_{\rm fold})$ where $E$ is the photon energy
          and $E_{\rm fold}$ is the folding energy, both measured in keV.}

\modes{power}{Power-law of the form $A_{\rm PL} E^{-\Gamma}$, where the 
  symbols have the same meaning as for the cutoffpl model.}

\modes{diskbb}{MCD model after \citey{mitsuda:84a}, $kT_{\rm in}$ is the
  temperature at the inner edge of the disk measured in keV ($T(r)\propto
  r^{-3/4}$ where $r$ is the radius) and the normalization constant is
  $A_{\rm diskbb}=(r_{\rm in}/d)^2 \cos i$ where $r_{\rm in}$ is the inner
  disk radius measured in kilometers, $d$ is the source distance, measured
  in units of 10\,kpc, and $i$ is the disk inclination.}

\modes{compTT}{Comptonization model after \citey{tit:94a} for a disk
  geometry. $kT_0$ is the temperature of the Wien like seed photon input
  spectrum measured in keV, $kT_{\rm e}$ is the plasma temperature, measured in
  keV, and $\tau$ is the plasma optical depth.}

\modes{ray}{Optically thin emission spectrum after \citey{raymond:77a}
  where $kT_{\rm ray}$ is the plasma temperature in keV. The normalization
  constant is $A_{\rm ray}=10^{-17} (4\pi D^2)^{-1} \cdot \int n_{\rm e}
  n_{\rm H} {\rm d}V$, integrating over the emitting volume, where $n_{\rm
    e}$ and $n_{\rm H}$ are the electron and hydrogen particle densities,
  respectively, and $D$ is the distance in cm.}

\tablecomments{The absorbing column, $N_{\rm H}$, has been fixed at
  $1.3\times 10^{21}\,{\rm cm^{-2}}$ for the PCA data, while it was fixed
  at the value found from the SIS data in the case of the GIS data. In all
  other cases it was a free parameter (measured in $10^{21}\,\rm cm^{-2}$).  The
  normalization of the SIS1 detector with respect to the SIS0 detector was
  $0.987\pm 0.006$, while that of the GIS3 with respect to the GIS2 was
  found to be $1.043\pm 0.004$ in all cases. The uncertainties given are at
  the 90\% level for one interesting parameter.}
\end{deluxetable}


\begin{figure*}

\centerline{\includegraphics[width=0.45\textwidth]{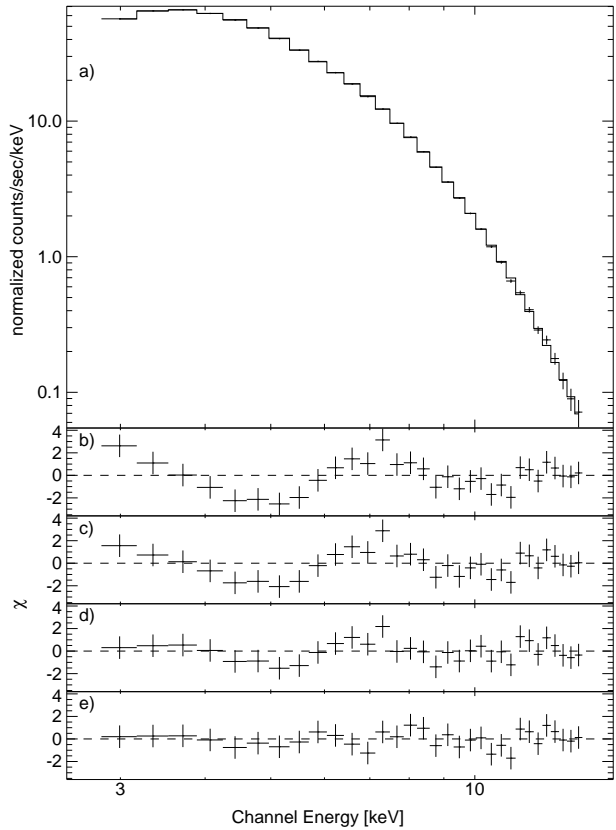}}

\caption{\small Spectral modeling of the PCA data. Residues are shown as the
  contribution to $\chi$. a) Count rate spectrum and the best fit
  MCD model with an optically thin plasma spectrum. b) Residues of the best fit
  exponentially cutoff power-law, c) residues of the best fit MCD model,
  d) residues of the best fit Comptonization
  model, and e) residues of the best fit MCD plus optically thin plasma
  spectrum.\label{fig:rxteplot}}

\end{figure*}

To describe the PCA spectrum we use the spectral models traditionally
applied to \aql: an exponentially cutoff power-law, a multicolor disk-black
body (\cite{mitsuda:84a}), and a Comptonization model after \citey{tit:94a}.
Due to the low sensitivity of the PCA to the low absorbing column towards
\aql\ (see \cite{stelzer:98a} for a discussion of the sensitivity of the
PCA to $N_{\rm H}$), we fixed $N_{\rm H}$ to the \citey{dickey:90a} value of
$N_{\rm H}=1.3\times 10^{21}\,{\rm cm^{-2}}$.  The results of our spectral
fits are given in Table~\ref{tab:fitres}, while the PCA spectrum and the
residues are displayed in Fig.~\ref{fig:rxteplot}.  All three models
roughly describe the observational data.  Note that we do not see any
evidence for a high energy power-law tail as that seen in previous
observations.  The 90\% confidence level upper limit to the 3--20\,keV flux
from a power law is $8\times10^{-12}~{\rm ergs~cm^{-2}~s^{-1}}$, which is
less than 2\% of the observed 3--20\,keV flux.

The best description of the PCA data is given by the Comptonization model
($\chi^2_{\rm red}=0.82$ for 29 degrees of freedom), while the residues of
the MCD model and the exponentially cut-off power-law show structure in
excess of that expected from calibration uncertainties of the PCA. These
residues are especially apparent in the low energy channels of the PCA,
below the characteristic feature of the Xe L-edge at $\sim 5$\,keV (a
region of very uncertain detector calibration; see the discussion by
\cite{wilms:98c}).  Inspection of our best fit values in
Table~\ref{tab:fitres} shows that the Comptonization model results in such
a good fit because the seed photon temperature of the model, taken here as
a Wien spectrum with best-fit temperature $kT_0=0.34$\,keV, is
uncharacteristically large. This is also evident in the very asymmetric
confidence contour indicated in Table~\ref{tab:fitres}.  Setting $kT_0$ to
a small value yields residues that more resemble those seen in the MCD and
exponentially cut-off models.  We conclude that there is unambiguous
evidence for the presence of an additional soft-excess below $\sim 5$\,keV.
Since the low-energy cut-off of the PCA is at $\sim 2$\,keV, this
instrument cannot be used to further constrain the nature of this
soft-excess. We therefore turned to archival data from observations of
\aql\ made with ASCA and ROSAT.

\begin{figure*}
\centerline{%
\includegraphics[width=0.45\textwidth]{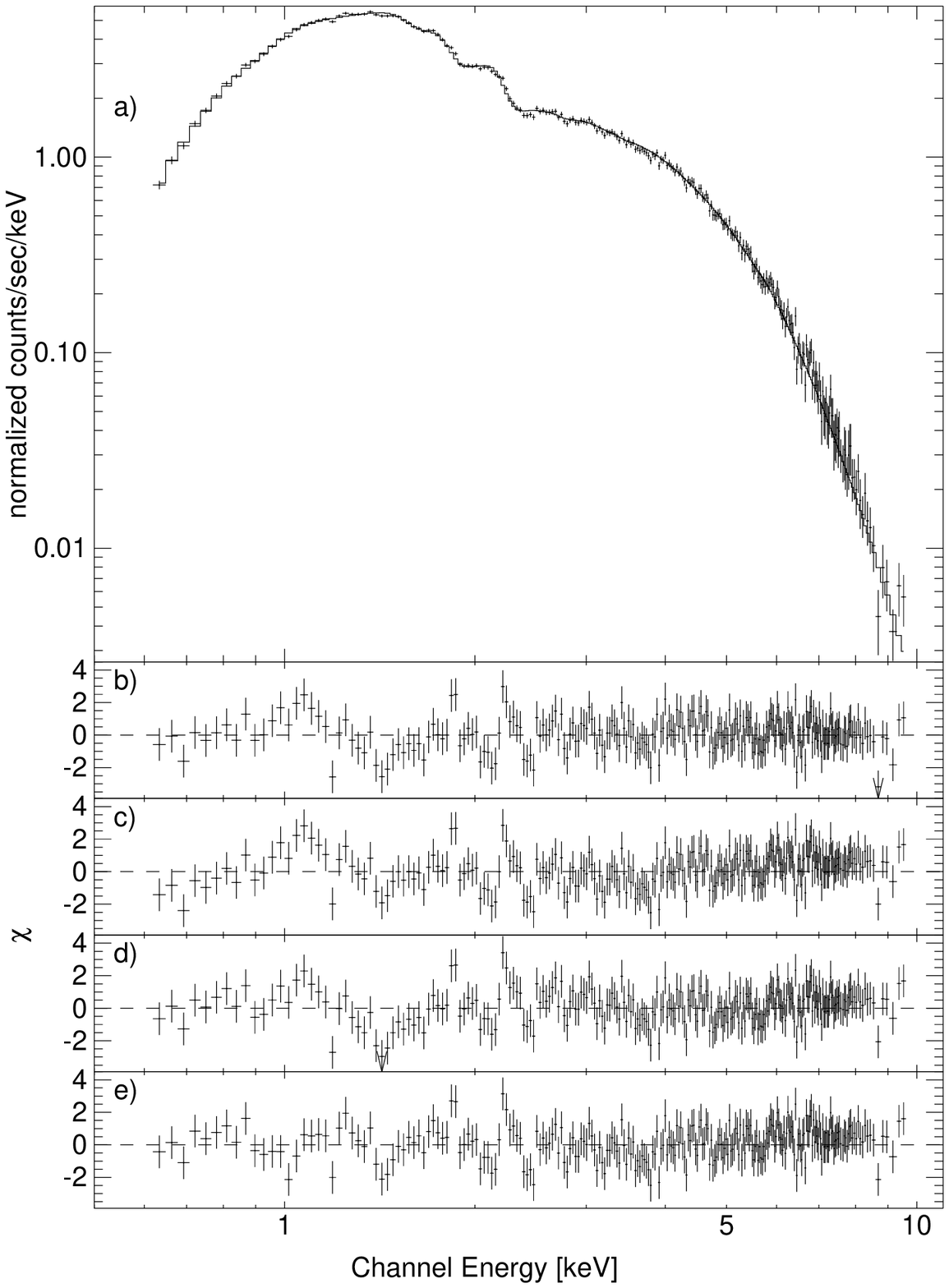}
\includegraphics[width=0.45\textwidth]{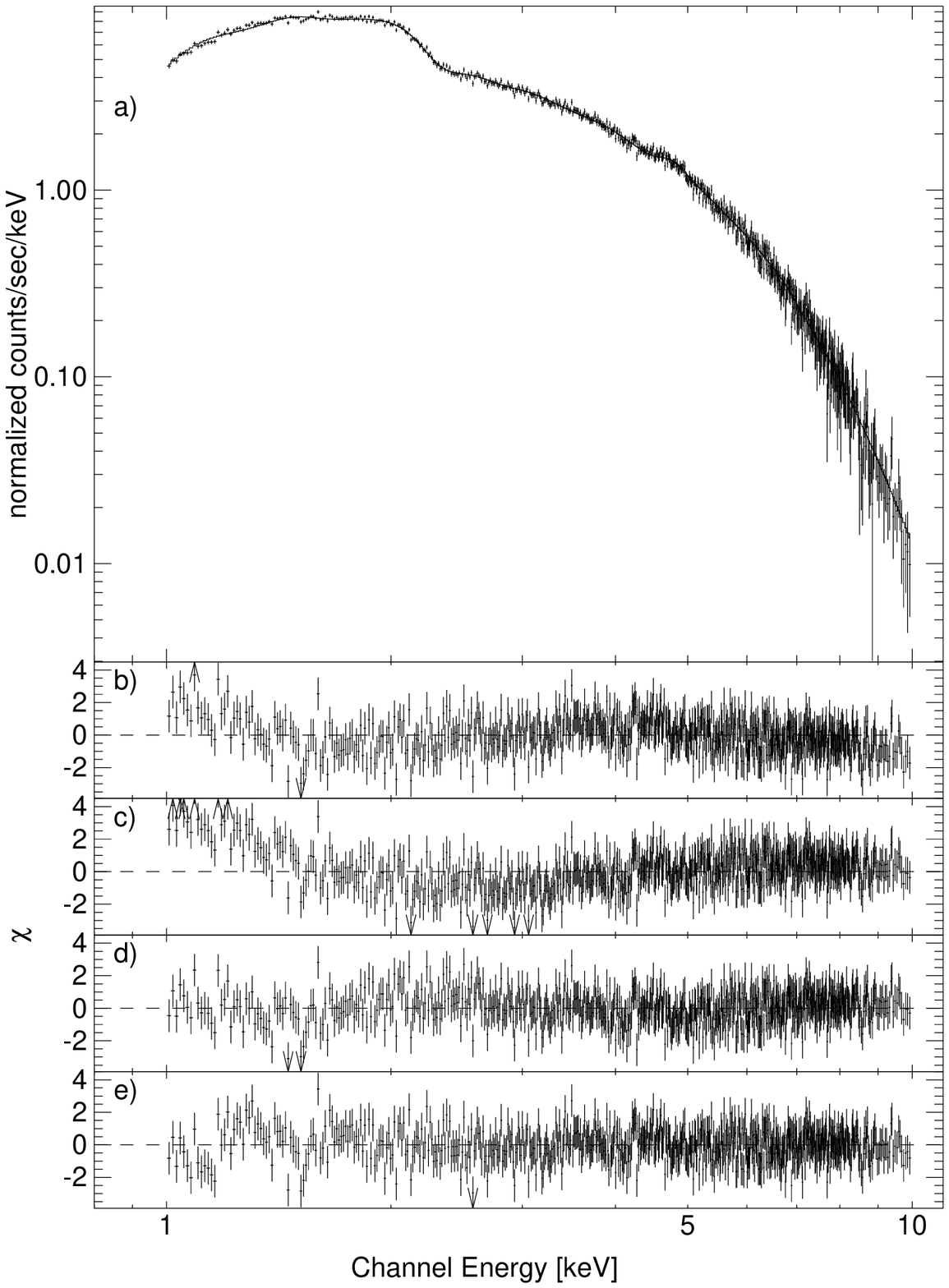}
}

\caption{\small Same as Fig.~\ref{fig:rxteplot} for the ASCA data.
  \emph{Left:} SIS data. \emph{Right: GIS data}. Residuals at 1.7 and 2.2
  keV are well-known ASCA response features. \label{fig:ascaplot}}
\end{figure*}

The High Energy Astrophysics Archive (HEASARC) contains one ASCA
observation of \aql, made in 1994~October (Table~\ref{tab:obslog}). A
preliminary analysis of these data has been presented by \citey{ricci:96a}.
We extracted the data from all four instruments on ASCA, the two solid
state detectors (SIS0 and SIS1) and the two GIS detectors (GIS2 and GIS3).
Due to the uncertainty in the intercalibration of the instruments, the GIS
and the SIS detectors were analyzed separately.  We describe the data
extraction process in appendix~\ref{sec:asca}.

\begin{figure*}

\centerline{\includegraphics[width=0.45\textwidth]{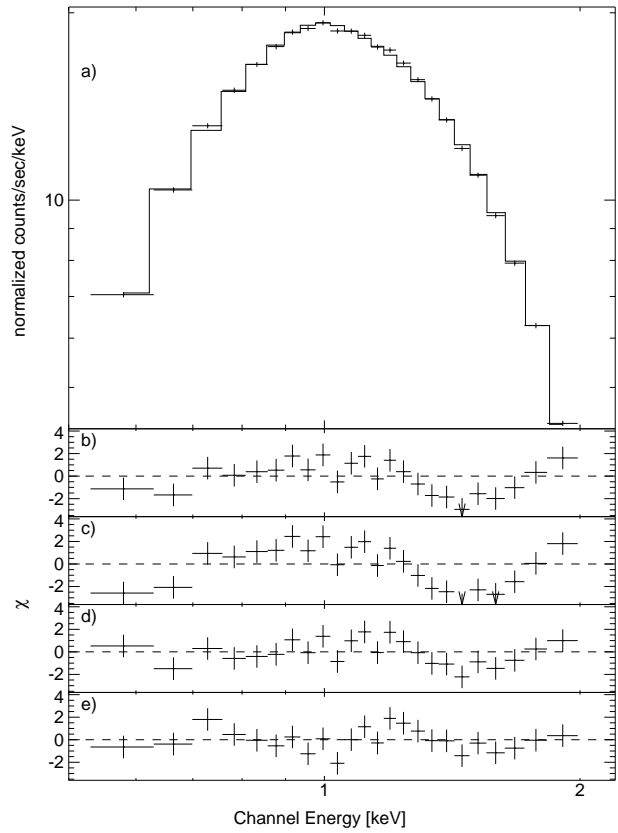}}

\caption{\small Same as Fig.~\ref{fig:rxteplot} for the ROSAT PSPC
  data.\label{fig:rosat}} 

\end{figure*}

In Table~\ref{tab:fitres} we list the results from modeling the data with
the standard models for the SIS and the GIS, respectively. The data and
residues for the models are shown in Fig.~\ref{fig:ascaplot}.  Note that
due to our extraction procedure the model normalizations differ between the
detectors. It is only possible to compare the spectral shapes (see
appendix~\ref{sec:asca}). As with the RXTE-PCA data, both the exponentially
cut-off power-law and the MCD model provide a rough description of the
data. Due to the higher spectral resolution of the ASCA detectors, however,
the causes for the spectral deviations are now apparent, and include a
strong deviation at $\sim 1$\,keV. We interpret this feature as evidence
for the presence of line emission at this energy, which might come from the
iron L complex or emission features from other metals such as K$\alpha$
lines from highly ionized neon or magnesium (see \cite{nagase:94a}).
Modeling the feature with the addition of a simple Gaussian line does not
result in a markedly improved fit.  In contrast, including an optically
thin thermal plasma spectrum after \citey{raymond:77a} in the spectral
modeling process results in a dramatic improvement of the fit ($\Delta
\chi^2_{\rm red}=0.65$ for the MCD model).  The best-fit parameters for the
thermal plasma are similar to the disk temperature found with the MCD model
and the emission line spectrum is dominated by emission around 1\,keV.

In order to further check whether the 1\,keV feature is always present in
the X-ray spectrum of \aql\ we turned to archival ROSAT position
sensitive proportional counter (PSPC) data. The
observing log for this observation is given in Table~\ref{tab:obslog} and
the data extraction procedure is described in appendix~\ref{sec:rosat}. As
can be seen from our fit-results in Table~\ref{tab:fitres}, the ROSAT data
give similar results as the ASCA data. In fact, the ROSAT data
\emph{require} the presence of the line emission component to provide
satisfactory fits (see also~Fig.~\ref{fig:rosat}).

We note that the PCA data shows weak residuals in the region of an Fe line.
An MCD model with weak power law tail, for example, admits the inclusion of
a 6.6\,keV line with width 0.8\,keV and equivalent width 80\,eV.  Such a
weak, broad line, however, is comparable to the remaining uncertainties in
the PCA response matrix, and we therefore cannot be confident of its
significance nor of its parameters.  Adding an Fe line (with energies
ranging from 6.4 to 7.1\,keV) to the models of the ASCA data also does not
significantly improve the fits.  Limits to the equivalent width of any line
in this region were of ${\cal O}(10\,{\rm eV})$, which is comparable to the
equivalent width of the Fe line present in the best fit Raymond-Smith
models.  We note that contrary to the EXOSAT and \textsl{Ginga} data, the
ASCA data also do not show strong evidence of a hard tail.  The upper limit
to the flux from a power law tail was 12\% of the 2--10\,keV flux in the
cutoff power law model of the ASCA GIS data.  The upper limits for the MCD
models and for the SIS models were 3--20 times lower.  We therefore cannot
rule out the possibility that the 7.06\,keV line claimed by
\citey{ricci:95a} was associated with the presence of a hard tail.

\section{Timing Analysis}\label{sec:temporal}

\begin{figure*}
  \centerline{\includegraphics[width=0.45\textwidth]{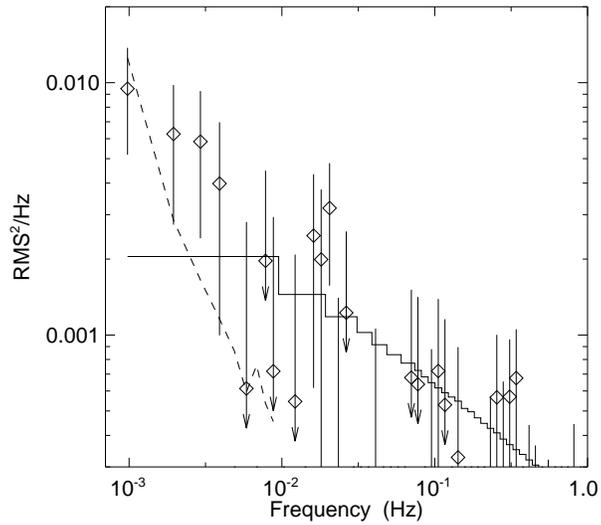}} 
\caption{\small Power spectral density for 1--7.2\,keV, logarithmically
  binned in frequency over $f \rightarrow 1.08 f$, and normalized as in
  \protect\cite{belloni:90b} and \protect\cite{miyamoto:92a}.  The solid
  line is the effective Poisson noise level for noise-subtracted data (see
  \protect\cite{vanderklis:89b,nowak:98a}). The dashed line is our estimate
  of the contribution from RXTE background fluctuations.
  \label{fig:psd}}
\end{figure*}

We employed Fourier techniques, in the same manner as for our RXTE
observations of Cyg X-1 (\cite{nowak:98a}) and GX~339$-$4
(\cite{nowak:99a}), to study the short timescale variability of \aql. We
use the same techniques for estimating deadtime corrections
(\cite{zhangw:95a,zhangw:96a}) to the Power Spectral Density (PSD), and for
estimating uncertainties and the Poisson noise levels of the PSD
(\cite{leahy:83a,vanderklis:89b}) as in our previous RXTE analyses.  We use
lightcurves with $2^{-5}$\,s resolution, from the PCA top layer data only,
over the energy range $\approx 1$--7.2\,keV (absolute PCA channels 1-20),
in the analysis discussed below.  We also searched $2^{-11}$\,s lightcurves
over the same energy range for high-frequency features, but none were found
above the Poisson noise limits.

As the source intensity did not appear to vary over the course of the
observation, we created a single PSD.  The results are presented in
Figure~\ref{fig:psd} for a normalization where integrating over positive
frequency yields the mean square variability (see
\cite{belloni:90b,miyamoto:92a}).  Note that above $f=10^{-2}$\,Hz the
power is completely consistent with Poisson noise. We estimate that the
background contributes 13\,cps to the lightcurves, compared to 210\,cps for
the signal.  Based upon these count rates and from calculating the PSD of
the background lightcurve generated using the RXTE software, we find that
the PSD observed between $f=10^{-3}$--$10^{-2}$\,Hz is consistent with
background fluctuations.  We find that the upper limit to the root mean
square (rms) variability between $10^{-3}$--$16$\,Hz is 4\%.

\section{Long-Term Variability}\label{sec:long}

\begin{figure*}
\centerline{
\includegraphics[width=0.45\textwidth]{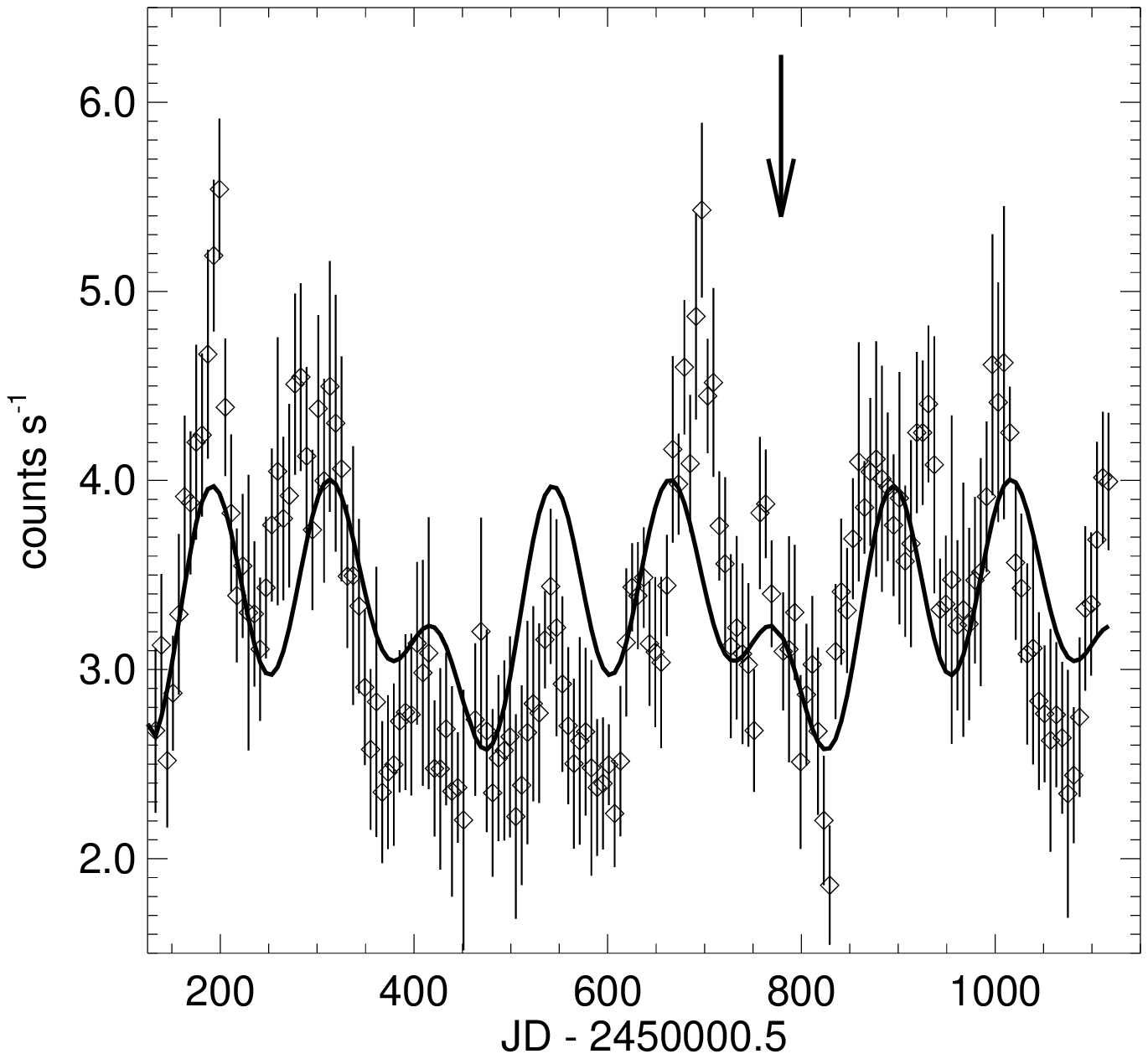}
\includegraphics[width=0.45\textwidth]{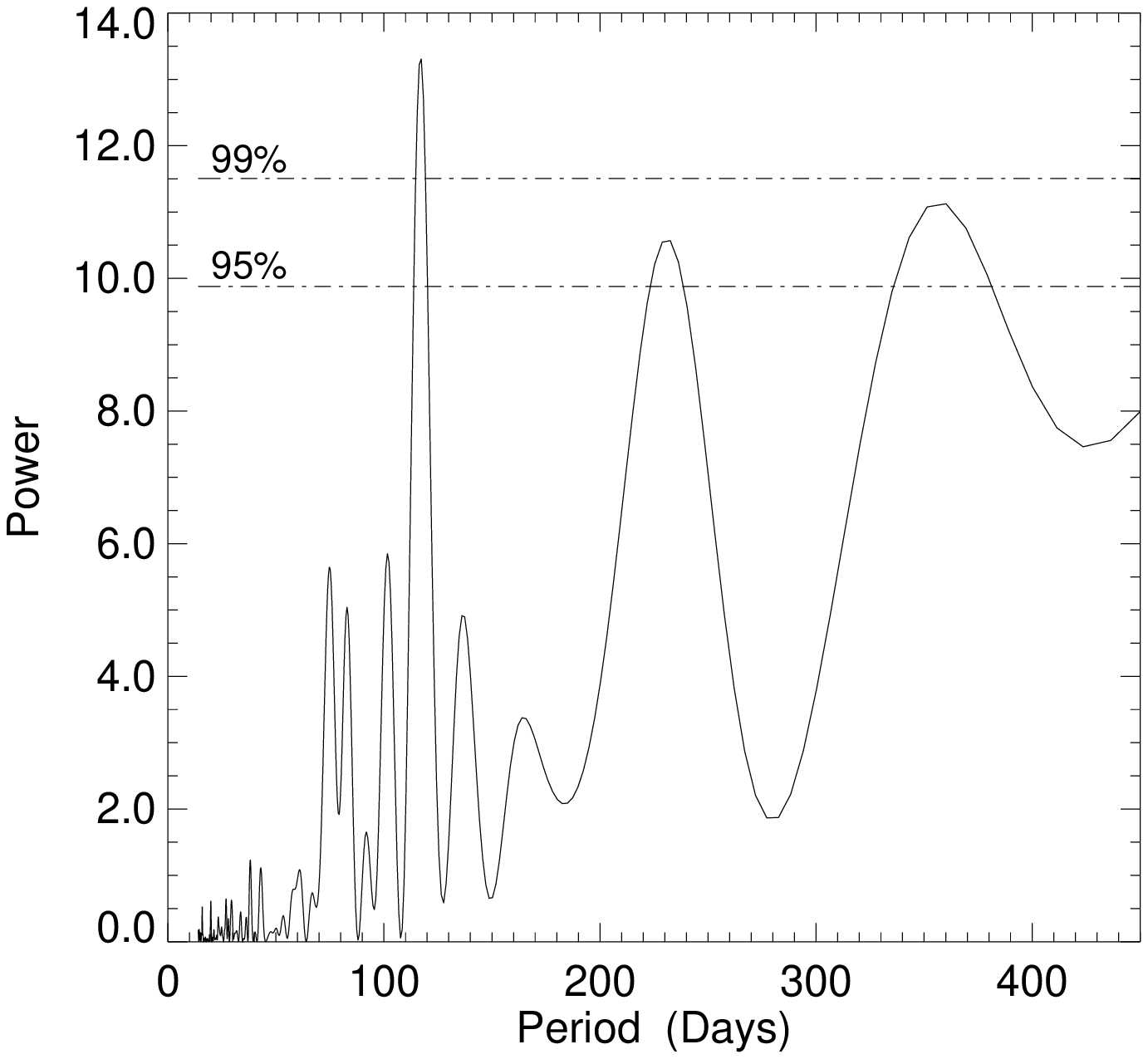}
}
\centerline{
\includegraphics[width=0.45\textwidth]{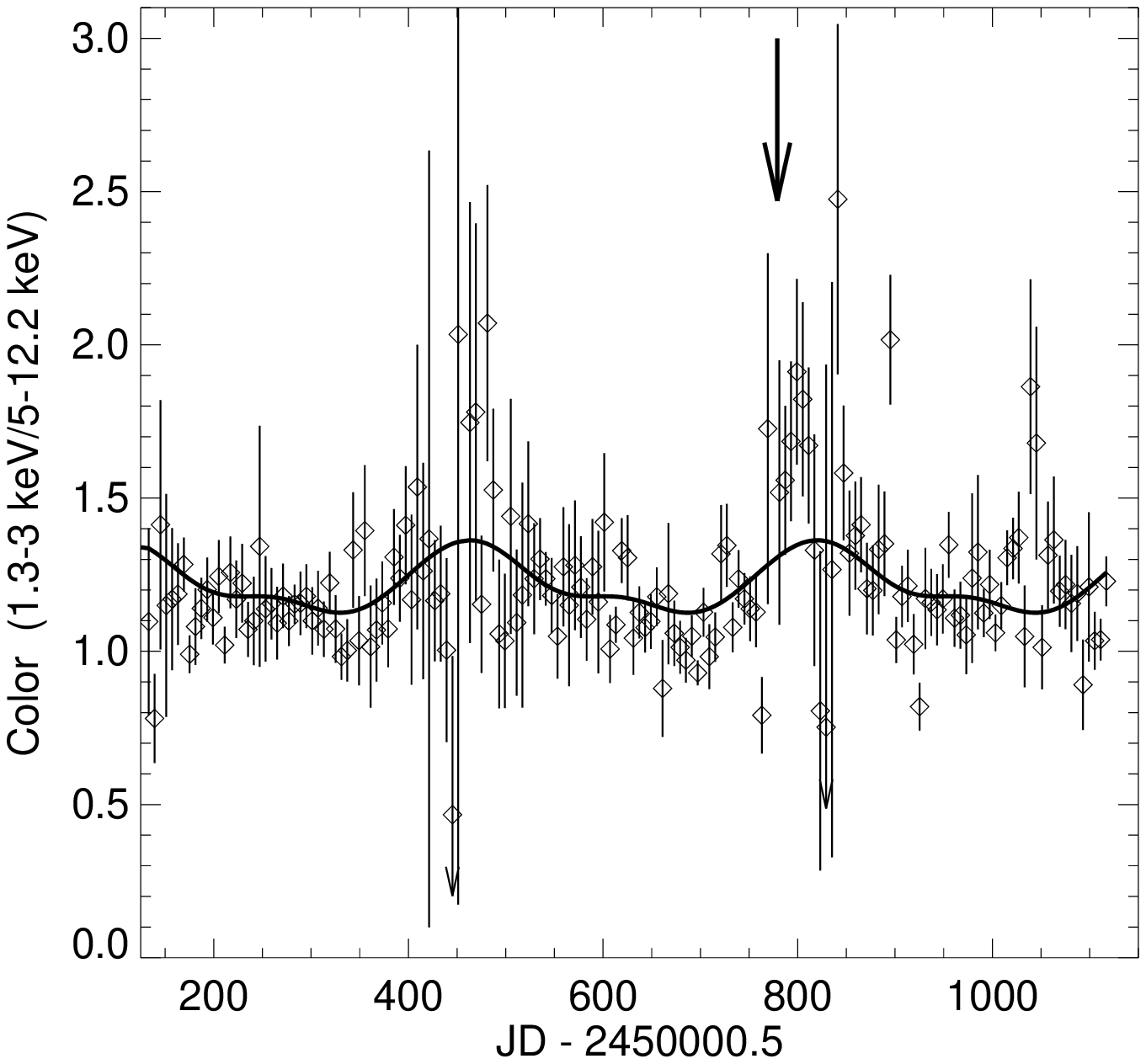}
\includegraphics[width=0.45\textwidth]{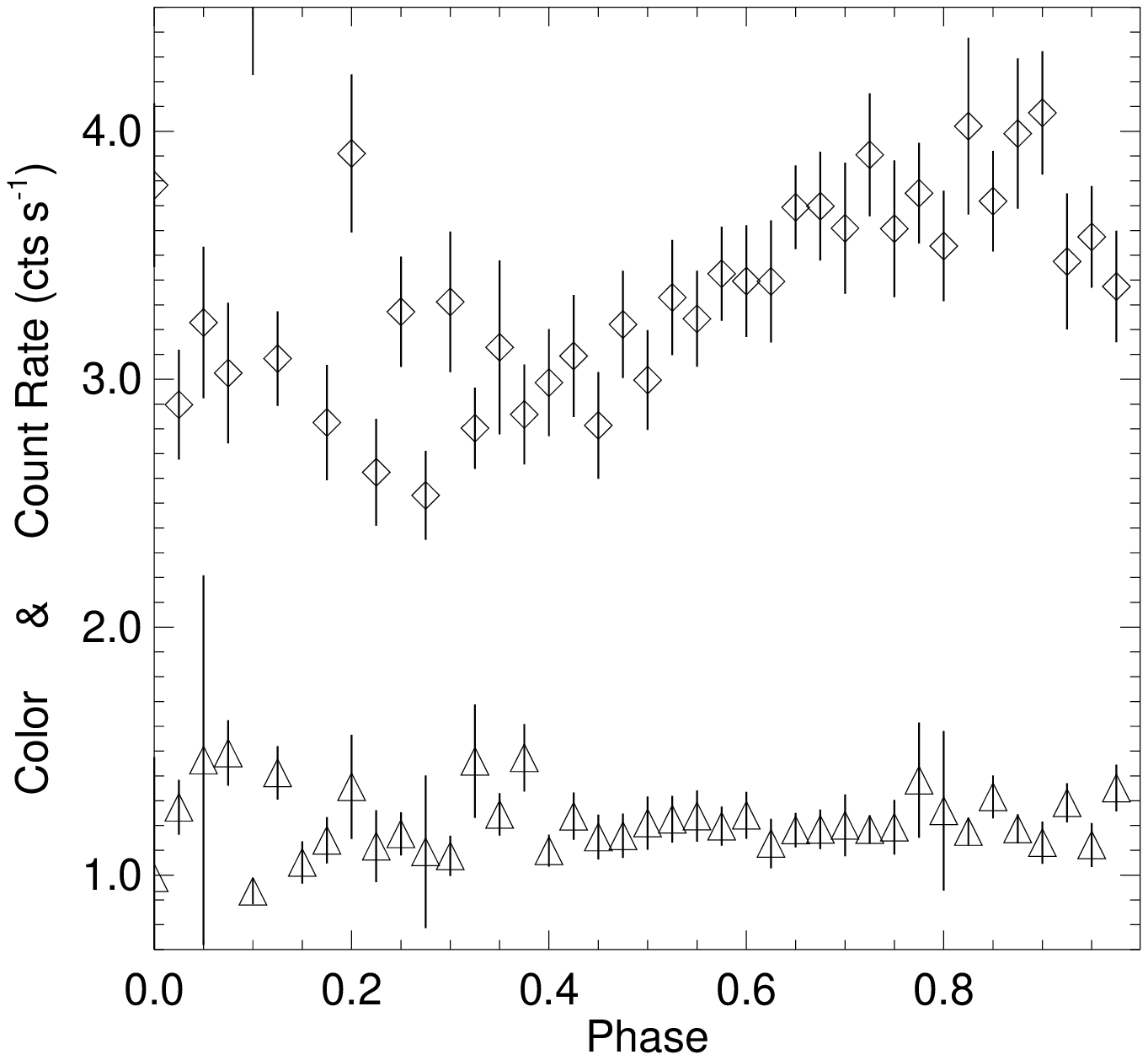}
}

\caption{\small {\it Top Left:} RXTE-ASM count rate lightcurve for
  V1408 Aql, 6 day binning, with sinusoids fit at periods of 117, 235, and
  352 days. {\it Bottom Left:} RXTE-ASM color lightcurve (1.3--3.3\,keV
  count rate divided by the 5--12.2\,keV count rate), 6~day binning, with
  sinusoids fit at periods of 117, 235, and~352 days.  {\it Top Right:}
  Lomb-Scargle Periodogram for RXTE-ASM count rate lightcurve binned at 6
  days. Lines are the 95\% and 99\% significance levels. {\it Bottom
    Right:} Count rate lightcurve (diamonds) and color lightcurve (ch.
  1/ch.  3; triangles) folded on a 117 day period. Arrows denote the date
  of our RXTE observation. \label{fig:asm}}
\end{figure*}

We used data from the All Sky Monitor (ASM) on RXTE to study the long-term
behavior of \aql. The ASM is an array of three shadow cameras combined
with position sensitive proportional counters that provides for a
quasi-continuous coverage of the sky visible from RXTE
(\cite{levine:96a,remillard:97a}).  Lightcurves in three energy bands ---
1.3--3.0\,keV, 3.0--5.0\,keV, and 
5.0--12.2\,keV --- as well as over the whole ASM band are publically
available from the ASM data archives (\cite{lochner:97a}).  Typically there
are several 90\,s measurements available for each day. 

In Figure~\ref{fig:asm} we present the ASM data of \aql\ that were
available as of 1998 November 20.  The date of our pointed RXTE observation
is indicated by an arrow in this figure.  Several features are immediately
apparent in these data.  The count rate light curve shows significant
variability with fluctuations up to ${\cal O}(50\%)$ of the mean.  These
fluctuations occur on ${\cal O}$(100 day) timescales.  The color lightcurve
(we show the 1.3-3.0\,keV lightcurve divided by the 5.0--12.2\,keV
lightcurve) shows significantly less variability, with peaks in the
softness of the source occurring on ${\cal O}$(400 day) timescales.
Furthermore, the peaks in the softness of the source seem to be correlated
with dips in the intensity of \aql. The features in the light curve appear
to be associated with possible long term periodicities.

We determined the significance of these possible long term periodicities by
computing the Lomb-Scargle Periodogram (\cite{lomb:76a,scargle:82a}) for
the 1.3--12.2\,keV band for 6-day averages of the ASM lightcurves. We
averaged data where the best fit to the source position and flux in an ASM
observation had a $\chi_{\rm red}^2 \le 1.5$ (see \cite{lochner:97a}) in
\emph{each} of the three ASM energy channels.  The periodogram presented in
Figure~\ref{fig:asm} shows evidence of a 117 day, 235 day, and a 352 day
periodicity. Each of these periodicities is significant at greater than the
95\% level, as determined by the methods of \citey{horne:86a}. We note that
the Lomb-Scargle periodogram does not assume the presence of harmonics;
this is a result of the analysis.  Epoch folding (see
\cite{leahy:83a,schwarz:89a,davies:90a}) of the ASM lightcurves also shows
evidence of these periodicities, although each period has uncertainties of
approximately $\pm 10$\,days.  The evidence for a periodicity in the color
lightcurve is somewhat weaker. Only the longest period appears, with an
approximately 370 day period, and then only at the 50\% significance level
in a Lomb-Scargle periodogram.

Figure~\ref{fig:asm} shows the result of fitting three harmonically spaced
sinusoids to the count rate and color lightcurves.  In these fits, the
periods were constrained to be within a few days of the periods found in
the Lomb-Scargle periodogram of the count rate lightcurve; however, the
phases of the sinusoids were left completely free. For the count rate
lightcurve, the amplitudes of the sinusoids are 0.32\,cps, 0.26\,cps, and
0.41\,cps for the fundamental, first harmonic and second harmonic.  For the
color lightcurve, the respective amplitudes are 0.1, 0.05, and 0.01.
Furthermore, the phases of the sinusoids are displaced from those of the
count rate lightcurves by of ${\cal O}(\pi)$.  In Figure~\ref{fig:asm} we
also show the lightcurves folded on the 117 day period.  Note that the
folded color lightcurve indeed exhibits very little variation on this
timescale.  The count rate lightcurve shows significantly more periodic
structure.  The low flux points, however, display the most variations from
phase bin to phase bin.  Partly this could be due to interference from the
235 and 352 day periods.  Additionally, if this periodicity is due to
inclination effects in a warped disk, as we further discuss below, the low
flux points come at times when the disk is at its greatest inclination to
our line of sight.  The lightcurve is most sensitive at these times to
small changes in disk thickness and/or shape.

Long timescale periodicities and quasi-periodicities are relatively common
in ASM observations of binary sources (Remillard 1997, private
communication).  Evidence for a 294\,d periodicity in Cyg~X-1 has been
previously reported (\cite{kemp:83a,priedhorsky:83a}), and is readily
apparent in the ASM data during the hard state.  A 198\,day periodicity
also has been observed in LMC~X-3 (\cite{cowley:91a,wilms:98d}), and a
possible 240 day periodicity appears in ASM data of the low/hard state of
GX~339$-$4 (\cite{nowak:99a}).

\section{Discussion --- The nature of V1408~Aql}\label{sec:disc}

To summarize, our spectral analysis has provided evidence for a very soft
spectrum which can be satisfactorily modeled with any of the three
traditional models used here, namely the exponentially cutoff power-law,
the MCD model, and Comptonization. We did not see any evidence for a hard
power-law tail similar to that seen in previous observations.  We have also
found evidence for a spectral feature at $\sim 1$\,keV, which we interpret
as emission from the iron L complex or as K$\alpha$ lines from highly
ionized metals. No short term variability in excess of the noise was
detected from the source, and the upper limit to the rms variability
between $10^{-3}$--$16$\,Hz is 4\%.  On long timescales, we found evidence
for periodic variability on a time-scale of about 117\,days in the soft
X-ray luminosity, and evidence for a periodic softening of the X-ray
spectrum on a 350--400 day timescale.  Below, we discuss interpretations of
these results.

\subsection*{Spectral Considerations}

Although the Comptonization model appears to provide the best fit to our
broad-band RXTE data, we do not regard it likely that Comptonization is
indeed the physical process responsible for producing the X-ray spectrum.
As we have shown in \S\ref{sec:spectral}, part of the small $\chi^2_{\rm
  red}$ obtained for Comptonization is attributable to the comparably high
seed photon temperature, $kT_0 = 0.34$\,keV, which mimics the soft excess
seen in the RXTE data. Also, the best fit parameters hint at a very cold
and optically thick Comptonizing plasma with an optical depth of almost 10.
Commonly assumed models for Compton coronae, such as advection dominated
accretion flows (\cite{esin:97c}) or other `sphere plus disk' coronal
models (\cite{dove:97b}), have considered only hot, optically thin to
moderately optically thick coronae.  It is not clear whether a cool and
very optically thick corona can be made energetically self-consistent, nor
is it clear what physical processes would lead to such a configuration.  We
therefore conclude that Comptonization is an improbable physical mechanism
for producing the observed soft spectrum.

The accretion disk spectrum and the exponentially cutoff power-law both
provided similar quality fits and yielded almost indistinguishable
residues.  The MCD model, however, seems the better phenemonological
representation of the underlying physical mechanism for producing the
observed spectrum.  The best fit parameters for the exponentially cutoff
power-law span a wide range (including a negative photon index in the RXTE
spectrum), and in many ways appear to be ``mimicing'' the features of the
MCD model. The MCD models, on the other hand, have best-fit spectral
parameters that are all similar for each of the independent observations.
More importantly, optical and infrared observations
(\cite{cowley:88a,shahbaz:96a,hakala:99a}) provide independent evidence for
the presence of an extended accretion disk in \aql.  As we discuss further
below, additional independent evidence for the assumption that the X-rays
are dominated by the accretion disk comes from the presence of the long
term spectral variability.

We note that the line features apparent in the ASCA and ROSAT data are also
consistent with an accretion disk picture. Line features around 1\,keV are
a common occurrence in photoionized plasmas close to sources emitting hard
X-rays (e.g., in eclipse in Vela~X-1, \cite{nagase:94a}). We would also
expect such features in models with warped accretion disks similar to those
of \citey{schandl:96a} (see discussion below). Iron~L features and
K$\alpha$ lines from Mg and Ne are also predicted in models for reflection
off ionized accretion disks (\cite{ross:93a}), and are in fact observed in
several NS-LMXB such as Cygnus X-2 (\cite{vrtilek:86a,kallman:89a}), albeit
the complexity of the observed line shapes makes a direct comparison
between the data and the models difficult. See \citey{kallman:96a} for a
detailed discussion of these features.

\subsection*{Long Term Variability}

The timescales of the periodicities observed with the ASM are comparable to
the timescales expected from precessing accretion disk warps, whether they
are driven by the radiation pressure instability discovered by
\citey{pringle:96a} (see also \cite{maloney:96a,maloney:97a,maloney:98a}),
or by an X-ray heated wind as for models of Her X-1 (\cite{schandl:96a}).
As radiation pressure must typically strongly dominate gas pressure before
a wind can be launched, the former mechanism may dominate
(\cite{maloney:97a}), at least for warps large enough such that the outer
disk is effectively lluminated by the X-ray flux from the inner disk.  This
radiation pressure driven instability is fairly generic, and is expected to
cause a radiatively efficient (i.e., non-advection dominated) accretion
disk to warp and precess on ${\cal O}(100~{\rm day})$ timescales.  The
observed ratio between the precession period and the orbital period in
\aql\ is too long to be explained by a tidally forced precession of the
accretion disk (\cite{larwood:98a}).

In a warped disk scenario, the long term modulations could be due to a
combination of the flux varying as the cosine of the inclination angle, as
well as due to obscuration of the inner disk by the outer disk.  For the
former effect, we note that if the inclination of \aql\ is
$70^\circ$--75$^\circ$ as suggested by \citey{hakala:99a}, then relatively
modest inclination variations of $\pm 10^\circ$ can yield the observed
X-ray luminosity variations.  The softening of the spectrum observed on the
352 day timescale could be due to a warp periodically obscuring the central
regions of the accreting system, which would explain why we do not detect
the hard power-law tail seen by the previous observations.  In analogy to
other soft sources such as LMC X-3 (\cite{wilms:98d}), we can assume that
this tail is produced in a small and comparably cold accretion disk corona
close to the compact object which is then obscured by the precessing warp.
A prediction of this scenario, therefore, is that a long term monitoring
campaign with an instrument capable of detecting the hard power-law tail
(e.g., {\it RXTE} or {\it BeppoSAX}), will detect a \emph{periodic} change
in the flux level of the power law, including a periodic disappearance of
this tail.

\begin{figure*}
\centerline{\includegraphics[width=0.95\textwidth]{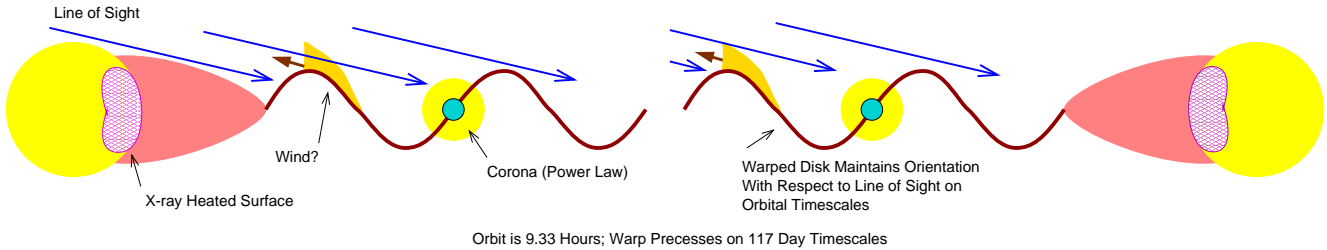}}
\caption{\small A possible model for the behavior of V1408~Aql (not to
  scale). Optical modulation is due to partial eclipsing of a warped disk.
  Long term X-ray variations are due to the changing orientation of the
  warp.  The X-ray power law is due to Comptonization within a corona.  The
  $\sim 1$\,keV plasma component is due to a wind or possibly due to a disk
  atmosphere.
  \label{fig:model}}
\end{figure*}

One alternative explanation is that the corona is covered by the rim of a
(geometrically thick) accretion disk. Unlike the warped disk scenario where
the relative inclination of the disk to our line of sight does not change
on orbital timescales (see Figure~\ref{fig:model}), in a disk rim scenario
the rim is caused by interaction of the accretion stream with the outer
edge of the disk.  Our relative view through the rim therefore changes on
orbital timescales (see \cite{hakala:99a} and references therein). This
seems to be less likely than the warp scenario, however, as contrary to the
optical and infrared data there appears to be no evidence for a modulation
of the X-ray spectrum on orbital timescales.  The warped disk picture could
also explain the observed change in the \emph{optical} lightcurve recently
discovered by \citey{hakala:99a} as a precession of a warp on long
timescales.

\subsection*{Short Term X-ray Variability}

Although we only have upper limits for the amplitude of the
$10^{-3}$--$16$\,Hz variability, these limits are consistent with the few
observations of BHC and some NS-LMXB in nearly ``pure'' soft states
(\cite{miyamoto:94b}).  For examples of BHC high/soft states, with little
or no discernible hard tail wherein short term variability is presented,
see \citey{grebenev:91a} for an observation of GX~339$-$4,
\citey{treves:88a} for an observation of LMC~X-3, \citey{ebisawa:89a} for
an observation of LMC~X-1, and \citey{miyamoto:94a} for observations of
Nova Muscae.  The PSD presented in these works typically have a PSD level
of $(\rm rms)^2/{\rm Hz}\approx 10^{-3}$ at 0.01\,Hz, which decreases as
$f^{-0.7}$ for higher Fourier frequencies and an rms variability of
$\approx 3\%$ in the $10^{-2}$--$30$\,Hz range.  This is slightly below the
upper limits presented in Figure~\ref{fig:psd}.

The `normal branch' of the NS-LMXB GX5+1 also has a similar amplitude and
shape PSD as described above for high/soft state BHC; however, its energy
spectrum consists of both a 1\,keV MCD component and a 2\,keV blackbody
spectral component (\cite{miyamoto:94b}, and references therein).  The
high/soft state of Cir~X-1 has been similarly modeled (\cite{miyamoto:94b},
and references therein). If \aql\ had weak 1--10\,Hz variability comparable
to that discussed by \citey{miyamoto:94b} for Cir~X-1, our observations
would have detected it.  Other soft neutron star sources with luminosities
of ${\cal O}(10\%)$ Eddington (the approximate luminosity of \aql, if it
were a 1.4~M$_\odot$ neutron star given our hypothesis of a highly inclined
disk), especially the bright atoll sources such as {GX13+1}, {GX3+1},
{GX9+1}, and {GX9+9}, can also exhibit ``very low frequency noise'' with
approximately 5\% rms variability (\cite{hasinger:89a}).  The
$10^{-4}$--100\,s low amplitude variability in the light curves of these
sources has been interpreted as intermittent, slow nuclear burning on the
surface of the neutron star (\cite{bildsten:93a,bildsten:95a}).  Low
frequency variability at such a level is absent in \aql.  Furthermore,
bright atoll sources often show a 0.1--10\,Hz power spectrum in excess of
the upper limits discussed here (\cite{hasinger:89a}). The level of the
0.1--10\,Hz PSD seen in GX13+1 (\cite{homan:98a}), for example, also would
have been easily detected in the PSD of \aql, yet was not.

\subsection*{Black hole or neutron star?}

The nature of the compact object as of now is not clear. The general
picture outlined above is similar to that seen in neutron star X-ray
binaries such as Sco~X-1, Cyg~X-2, and others. \citey{yaqoob:93b} pointed
out that the normalization of the best-fit MCD model appears to indicate
that the compact object is a neutron star. This argument, however, strongly
relies on the assumed distance to \aql, for which no compelling measurement
exists (a lower limit of 2.5\,kpc comes from the fact that the ASCA and
ROSAT measured $N_{\rm H}$ values are consistent with the full galactic
column), and also relies on the assumption that the accretion disk is seen
closer to \emph{face on}.  The recent optical and soft X-ray
variability measurements, however, make a large inclination more probable.

Taking these points into account and assuming for the sake of argument a
source distance of 7\,kpc, then the overall flux of \aql\ is comparable to
that of the high state of the black hole candidate GX~339$-$4, which is a
very plausible BHC. The upper limits to the high frequency variability
discussed above are consistent with previously observed BHC power spectra
in high/soft states.  If transitions from the hard state to the soft state
occur at 5\%-10\% of the Eddington luminosity (see \cite{nowak:95a}), then
the compact object in \aql\ is consistent with being a 2--3~${\rm M}_\odot$
black hole.  Thus, although there is no compelling evidence that \aql\ 
contains a black hole, there also is no compelling evidence that \aql\ is a
neutron star.

\subsection*{Conclusions} 

X-ray spectroscopy and the study of both the long term and the
short term variability of \aql\ make a system geometry as that depicted in
Figure~\ref{fig:model} seem likely. A low-mass main sequence star serves, via
Roche Lobe overflow, as the donor for a compact object which is surrounded
by a large accretion disk which in turn dominates the system at all
wavelength ranges. The accretion disk is surrounded by an optically thin
plasma, either in the form of an accretion disk wind or a stationary
accretion disk photosphere, which emits the observed X-ray line radiation.
A small hot corona directly surrounding the compact object produces the
hard X-ray power-law. The whole accretion disk precesses on a time scale of
about 117\,d, obscuring the central region and causing the power-law tail
to periodically disappear and reappear. Also on these long timescales, the
changing view of the warp causes the orbital modulation of the optical
light-curve (due to partial obscuration of the outer accretion disk) to
vary from sinusoidal (\cite{thorstensen:87a}) to a more complex pattern
(\cite{hakala:99a}).

The nature of the compact object in \aql\ is still ambiguous.  We have put
forth a hypothesis, however, that might explain the observed phenomenology
and makes predictions that are observationally testable.  X-ray monitoring
over the 117\,d period with an instrument like {\it RXTE} or {\it BeppoSAX}
should reveal whether the X-ray power-law tail really does periodically
disappear and reappear as predicted by our model.  Furthermore, if the
source is at 10\% $L_{\rm Edd}$ and contains a neutron star, then about one
``Type I'' microburst per day might be expected (\cite{bildsten:95a}).
This should be easily observable during such a campaign.  One might also
hope to find ``kilohertz QPO'' (\cite{vanderklis:98a}), as are often
associated with atoll sources.  For these latter two possibilities,
however, we note that some of the brighter atoll sources such as GX13+1
have yet to exhibit kilohertz QPO (\cite{homan:98a}, and references
therein), and rarely exhibit Type I bursts (see, for example,
\cite{matsuba:95a}, and references therein).  Finally, high spectral
resolution observations as will be provided by the upcoming new generation
of X-ray instruments, such as the gratings on the Advanced X-ray Astronomy
Facility (AXAF) and the X-ray Multiple Mirror Mission (XMM), will provide
the spectral resolution necessary for resolving and studying the Fe L
complex. This will allow the application of plasma spectroscopic
diagnostics (e.g., \cite{liedahl:92a}) to the study of this fascinating
source.

\acknowledgements We thank Neil Brandt and Christopher Reynolds for
valuable advice concerning the ASCA data analysis.  We would also like to
acknowledge useful correspondence with Lars Bildsten and Rob Fender.  Ingo
Kreykenbohm made some literature references available to us. We thank Erik
Kuulkers for pointing out to us additional references. Rudy Wijnands
provided invaluable advice concerning the timing analysis.  This work has
been financed by NASA Grants \mbox{NAG5-3225}, \mbox{NAG5-4737}, and
\mbox{NAG5-7024}. JW was also supported by a travel grant from the
Deutscher Akademischer Austauschdienst. This research has made use of data
obtained through the High Energy Astrophysics Science Archive Research
Center Online Service, provided by the NASA/Goddard Space Flight Center,
and the ROSAT archive at the Max Planck Institut f\"ur Extraterrestrische
Physik in Garching bei M\"unchen.

\appendix

\section{Data Analysis Methodology}
\subsection{RXTE Data Analysis}\label{sec:rxte}
Our RXTE data were analyzed using the same procedure as that for our
analysis of the spectrum of GX~339$-$4 (\cite{wilms:98c}). Screening
criteria for the selection of good on-source data were that the source
elevation was larger than 10$^\circ$. Data measured within 30\,minutes of
passages of the South Atlantic Anomaly or during times of high particle
background (as expressed by the ``electron ratio'' being greater than 0.1)
were ignored. Using these selection criteria, a total exposure time of
27\,ksec was obtained.  To increase the signal to noise level of the data,
we restricted the analysis to the first anode layer of the proportional
counter units (PCUs) where most source photons are detected (the particle
background is almost independent of the anode layer), and we combined the
data from all five PCUs.

To take into account the calibration uncertainty of the PCA we applied the
channel dependent systematic uncertainties described by \citey{wilms:98c}.
These uncertainties were determined from a power-law fit to an observation
of the Crab nebula and pulsar taking into account all anode chains;
however, they do also provide a good estimate for the first anode layer
only since most of the photons are detected in this layer.

Since \aql\ has a comparably small count rate we are able to use the new
background model for the PCA that was made available by the RXTE Guest
Observers Facility (GOF) in 1998~June. The quality of this model was
checked by looking at high detector channels which are completely
background dominated. Although the measured count rate of \aql\ was at the
high end of the applicability of the new background model, the agreement
between the model and the measured background was good. This is in part due
to the fact that \aql\ is a very soft source which allows greater latitude
in using the background model for faint sources. Remaining background
residuals were minimized by using the XSPEC ``corfile'' facility which
renormalizes the background flux to decrease the best fit $\chi^2_{\rm
  red}$. The corrections applied to the background flux were on the order
of $\lesssim$1.5\%, indicating that at least for this source the background
model provides a good background estimate. Since the spectrum is completely
background dominated above 20\,keV, and due to the calibration uncertainty
below 3\,keV, we restricted the spectral analysis to the range from 3 to
20\,keV.

For the timing analysis, we generated lightcurves from the `GoodXenon'
data. Note that although there are short data gaps of 1--4\,s duration that
are flagged by commensurate jumps in the value of the time coordinate from
one data bin to the next, there are occasional data gaps where the
extraction software generates a continuous series of time bins despite the
data losses.  These data gaps \emph{do not} appear in lightcurves generated
from the `standard2f' data (which is processed by a different event
analyzer on-board RXTE).  These gaps can be recognized, however, in the
high time resolution data by searching for any sequence, 1\,s or greater in
length, of time bins with zero count rate. Four such `unflagged' sequences,
with 16\,s duration each, were found in our data.  (Aside from these four
16\,s sequences, there were a few instances where two $1/32$\,s time bins
in a row would have zero detected counts.  The lack of counts in these bins
were consistent with counting statistics, and we did not consider these to
be data gaps.)  The power spectra that we presented in Figure~\ref{fig:psd}
were made from continous data segments without internal data gaps.  If we
include data segments with the unflagged data gaps in the calculation of
the PSD, we obtain a low amplitude (5\% rms) PSD that is flat from
$10^{-3}$--$10^{-2}$\,Hz and is exponentially cutoff at higher Fourier
frequencies.  In fact, the presence of unflagged data gaps can be deduced
from such a characteristic PSD shape (Wijnands 1999, priv. comm.).

\subsection{ASCA Data Analysis}\label{sec:asca}

We extracted data from the two solid state detectors (SIS0, SIS1) and the
two gas detectors (GIS2, GIS3) onboard ASCA by using the standard ftools as
described in the ASCA Data Reduction Guide (\cite{day:98a}).  The data
extraction regions were limited by the fact that all the observations were
in 1-CCD mode and that the source was placed close to the chip edge. To
maximize the extraction regions, we chose rectangular regions of $6'\times
8'$ and $6'\times 7'$ for SIS0 and SIS1, respectively. 
Choosing a rectangular region does not effect the shape of the extracted
spectrum; however, the ASCAARF ancillary response matrix generator assumes
a circular region, so the flux normalization is slightly off (hence the
$\approx 30\%$ normalization differences between the SIS and GIS detectors
in Table~\ref{tab:fitres}).  For the GIS detectors
we chose circular regions centered on the source each with a radius of
$13'$.

The SIS count rate for \aql\ is large enough that the central regions of
the CCD suffer from pileup (i.e., two or more events being registered as a
single event).  Estimates of the amount of this pileup can be found in the
appendix presented by \citey{ebisawa:96b}.  Based upon our measured
spectrum and these estimates, we chose to exclude from analysis central
rectangular regions with dimensions of $4'\times 3'$ and $3'\times 3'$ for
SIS0 and SIS1, respectively.  With these exclusions, we estimate that
pileup will contribute less than 1\% of the counts at 10\,keV.

We used the SISCLEAN and GISCLEAN tools (\cite{day:98a}), with the default
values, to remove hot and flickering pixels. As the spectrum of \aql\ is
very similar to the low flux level of Cir~X-1 described by
\citey{brandt:96a}, we filtered the data with the same cleaning criteria
outlined in that work; however, we took the slightly larger values of
$10^\circ$ for the minimum elevation angle and 7\,$\mbox{GeV}/c$ for the
rigidity.  Also similar to the work of \citey{brandt:96a}, we formed
background estimates by extracting a circular region of radius $5'$
near the edge of the detector for the GIS observations.  For the SIS
observations, we chose L shaped regions near the corner of the chip
opposite from the source. Background, however, contributes relatively
little to the observations.

We rebinned the spectral files so that each energy bin contained a minimum
of 20\,photons. We retained SIS data in the 0.6 to 10\,keV range and GIS
data in the 1 to 10\,keV range. The cross-calibration uncertainties among the
instruments were accounted for by introducing a multiplicative
constant for each detector in all of our fits.  As discussed above, the
resulting data files showed reasonable agreement between all four
detectors.

\subsection{ROSAT Data Analysis}\label{sec:rosat}

The extraction of the ROSAT spectrum was performed using the standard ROSAT
PSPC data analysis package Extended X-ray Scientific Analysis System
(EXSAS) (\cite{zimmermann:98a}) following the procedures described by
\citey{brunner:97a}. Source counts were extracted from a circular region
centered on the position of \aql\ with a radius of $2'$, while the background
was extracted from an annulus centered on the source from which source
counts from detected background sources were removed. A correction for the
telescope vignetting was applied to the standard ROSAT response matrix. The
spectrum was then rebinned into 26 channels of $\sim 10000$\,counts each to
ensure an even signal to noise ratio over the whole ROSAT energy band. As
for RXTE and ASCA, the spectral analysis of the extracted data was then
performed with XSPEC, ignoring data measured below 0.5\,keV and above
2.5\,keV. 


\end{document}